\documentstyle[12pt]{article}

\begin{document}

\begin{titlepage}

\title{
\begin{flushright}
{\normalsize EPHOU-93-002\\
November, 1993\\
\ \\
\ \\}
\end{flushright}
A Mean Field Analysis of One Dimensional
Quantum Liquid with Long Range Interaction}
\author{Kenzo Ishikawa and Nobuki Maeda\\
{\it Department of Physics, Hokkaido University,}\\
{\it Sapporo, 060 Japan}}
\date{}

\maketitle

\begin{abstract}
\baselineskip=15pt

Bi-local mean field theory is applied to one dimensional
quantum liquid with long range $1/r^2$ interaction, which has exact
ground state wave function.
We obtain a mean field solution and an effective action which expresses a
long range dynamics. Based on them the ground state energy and correlation
functions are computed. The ground state energy agrees fairly well with the
exact value and exponents have weaker coupling constant dependence than that
of partly known exact value.
\end{abstract}

\end{titlepage}
\baselineskip=15pt

\newpage

Recently there are much interests in one dimensional quantum liquid
with $1/r^2$ potential. Ground state wave function and thermodynamic
quantities were found by Sutherland\cite{F} and Calogero\cite{G}.
Ground state wave function is of Jastrow type and becomes simple form
in certain values of coupling constant. It is possible to compute
exponents of correlation functions for these values of coupling constant.
For other general values it has been impossible to compute them.
Kawakami and Yang\cite{A} computed recently the exponents based on conformal
field theory. They show that the system belongs to Luttinger liquid
\cite{C}\cite{B}\cite{D}.

In the present paper we apply a bi-local mean field theory.
Mean field theory has been successfully applied to many places.
It is not clear if a mean field theory can be applied to strongly correlated
systems\cite{I}\cite{E}.
We apply such a mean field theory that includes correlations and compare
results with partly known exact values.
We see that a fair agreement is obtained in the ground state energy but an
agreement is not good for the exponents of correlation functions if the
linearized action is used.

Lagrangian which discribes interacting fermion system is
\begin{equation}
{\cal L}=\psi^\dagger (i{\partial\over\partial t}-eA_0)\psi
-\psi^\dagger {(\vec p+e\vec A)^2\over 2m}\psi-\int d\vec y
\psi^\dagger (x)\psi^\dagger (y)
{V(\vec x-\vec y)\over2}\psi(y)\psi(x),
\end{equation}
where $A_0$ and $\vec A$ are external electromagnetic potentials
and $V(x-y)$ is the two body potential energy between electrons.
We concentrate on the repulsive long range interaction,
$V(\vec x-\vec y)=g/\vert\vec x-\vec y\vert^2$.
Green's functions are computed from the partition function as
\begin{equation}
{\cal Z}=\int {\cal D}\psi^\dagger {\cal D}\psi e^
{-\int^\beta_0 d\tau\int dx {\cal L}_E},
\end{equation}
where $\beta=1/kT$ and $\tau=it$
in the path integral quantization method.
The fermionic fields $\psi(x)$ and $\psi^\dagger(x)$ are anti-commuting
c-numbers.
We write the interaction term as
\begin{equation}
-{1\over2}\psi^\dagger(x)\psi(y){V(x-y)\over2}\psi^\dagger(y)\psi(x)
(2-p(x-y))
\end{equation}
$$
+{1\over2}\psi^\dagger(x)\psi(x){V(x-y)\over2}
\psi^\dagger(y)\psi(y)p(x-y),
$$
where $p(x-y)$ is a c-number function and a suitable form is chosen
depending upon the interaction and dimension of space.
In the present paper we use $p(x-y)=1$.

Let us describe the system with one spatial dimension here, but its extension
is straightforward.
We rewrite the partition function\cite{H} using bi-local auxiliary field
$U(x_1,y_1;x_0)$ that is local in time coordinate and
bi-local in space coordinate.
\begin{equation}
{\cal Z}={1\over N}\int{\cal D}U
{\cal D}\psi^\dagger{\cal D}\psi e^{
-\int dx_0dx_1[{\cal L}_0+\int dy_1{\cal L}_{eff}]},
\end{equation}
$$
{\cal L}_0=\psi^\dagger(x)({\partial\over\partial x_0}+eA_0
)\psi(x)+\psi^\dagger(x){(p_1+eA_1)^2\over 2m}\psi(x),
\qquad\qquad\qquad\qquad\qquad
$$
\begin{equation}
{\cal L}_{eff}=[\{U(x_1,y_1;x_0)U(y_1,x_1;x_0)-U(x_1,y_1;x_0)
\psi^\dagger(y)\psi(x)
\qquad\qquad\qquad\qquad
\end{equation}
$$
-U(y_1,x_1;x_0)\psi^\dagger(x)\psi(y)\}
-\{U(x_1,x_1;x_0)U(y_1,y_1;x_0)
$$
$$
-U(x_1,x_1;x_0)
\psi^\dagger(y)\psi(y)
-U(y_1,y_1;x_0)\psi^\dagger(x)\psi(x)\}]
{1\over4}V(x-y),
$$
$$
N=\int{\cal D}U e^{-\int dx_0dx_1dy_1{V(x_1-y_1)\over 4}
\{U(x_1,y_1;x_0)U(y_1,x_1;x_0)
-U(x_1,x_1;x_0)U(y_1,y_1;x_0)
\} }.
$$
The partition function goes back to the Eq.(2) by integrating the
bi-local field $U(x_1,y_1;x_0)$ in Eq.(4). We write $\cal Z$,
by integrating the fermionic field first, as
\begin{equation}
{\cal Z}={1\over N}\int{\cal D}U e^{-S_{eff}(U)},
\end{equation}
$$
e^{-S_{eff}(U)}=\int{\cal D}\psi^\dagger{\cal D}\psi e^{
-\int dx_0dx_1[{\cal L}_0+\int dy_1{\cal L}_{eff}]}.
$$

{\it (A) Stationary phase approximation} ---
The integration of the bi-local field is defined around a
minimum of the classical action $S_{eff}(U)$,
{\it i.e.}, a mean field $U_0(x_1,y_1)$
which satisfies a self-consistency condition,
$$
{\partial S_{eff}(U)\over \partial U}\biggl\vert_{U=U_0}=0,
$$
\begin{equation}
\langle\psi^\dagger(x)\psi(y)\rangle_
{\lower1ex\hbox{$x_0=y_0\atop{U=U_0}$}}=U_0(x_1,y_1),
\end{equation}
in stationary phase approximation.
If $U_0(x_1,y_1)$ is a function of $x_1-y_1$, the system is
invariant under translation and momentum is a good quantum number.
The ground state in a mean field approximation is defined
as a state vector in which one particle states
up to Fermi momentum are filled. Hence the functions
$$
A_0=\mu,\ A_1=0,
$$
\begin{equation}
U_0(x_1,y_1)=\frac{\sin \pi r \rho}{\pi r},
\end{equation}
where $r=x_1-y_1$,
satisfy the self-consistency condition.
$\rho$ is the density and is connected with the Fermi momentum by
$p_F=\pi \rho$ and $\mu$ is the chemical potential.
Using the above $U_0(x_1,y_1)$ of Eq.(8),
$\cal Z$ is written as ${\cal Z}={\cal Z}_0\tilde{\cal Z}$
\begin{equation}
{\cal Z}_0=e^{-S_{eff}(U_0)},
\end{equation}
$$
\tilde{\cal Z}=\frac{1}{N{\cal Z}_0}\int{\cal D}U
{\cal D}\psi^\dagger{\cal D}\psi e^{-S(U_0+U,\psi^\dagger,\psi)},
$$
$$
S(U,\psi^\dagger,\psi)=\int dx_0dx_1({\cal L}_0+
\int dy_1 {\cal L}_{eff}).
$$
The mean field part, ${\cal Z}_0$,
corresponds to the mean field Hamiltonian,
\begin{equation}
H_m=\int dx_1[\psi^\dagger(x){p_1^2\over 2m}\psi(x)+
\psi^\dagger(x)F(p_1)\psi(x)]+E_{background},
\end{equation}
$$
F(p_1)=-2\int dr\frac{V(r)}{4}\{U_0(r)
e^{ip_1r}-U_0(0)\},
$$
$$
E_{background}=\int dx_1dy_1\frac{V(x-y)}{4}\{
\vert U_0(x_1-y_1)\vert^2
-U_0(x,x)U_0(y,y)\}.
$$
The second term of the kinetic energy and the background energy
are generated from the mean field part
and have the following forms,
$$
F(p)=\left\{
\begin{array}{l r}
g(p^2+p_F^2)/4,&\qquad\vert p\vert<p_F,\\
\qquad g p_F \vert p\vert/2,&\qquad \vert p\vert >p_F,\\
\end{array}\right.
$$
\begin{equation}
E_{background}=-\int dx_1\frac{g}{6}\frac{p_F^3}{\pi}.
\end{equation}

In mean field approximation, in which
we take only ${\cal Z}_0$,
the energy density, the fermion propagator and
the density correlation function become as
\begin{equation}
{\cal E}_{mean\ field}=\int^{p_F}_{-p_F}\frac{dp}{2\pi}
\{\frac{p^2}{2m}+F(p)\}-\frac{g}{6}\frac{p_F^3}{\pi}
=\frac{1}{6\pi m}p_F^3+\frac{g}{6}\frac{p_F^3}{\pi},
\end{equation}
\begin{equation}
\langle\psi^\dagger(x_1,x_0)\psi(y_1,x_0)\rangle
=\frac{\sin p_F r}{\pi r},\qquad\qquad
\end{equation}
$$
\langle \rho(x_1,x_0)\rho(y_1,x_0)\rangle=
\frac{1}{2\pi^2}\frac{1}{r^2}-\frac{1}{2\pi^2}\frac{\cos 2p_F r}{r^2}.
$$
Obviously the self-consistency condition Eq.(7) is satisfied.

$\tilde{\cal Z}$ in Eq.(9) is due to the fluctuations
around the mean field, and gives a correction to the above
energy density and propagator.
We compute their effects in the following using
local field expansions of the bi-local field.
The bi-local field is expanded by a complete set of
relative coordinates as
\begin{equation}
U(x_1,y_1;x_0)=U_0(x_1-y_1)[1+
\sum_{n=0}^{\infty}q_n(x_1-y_1)a_n(\frac{x_1+y_1}{2},x_0)],
\end{equation}
$$
\int dr\frac{V(r)}{4}[\vert U_0(r)\vert^2q_n(r)
q_m(-r)-\vert U_0(0)\vert^2q_n(0)q_m(0)]=\delta_{n,m},
$$
$$
q_0=iC_0,\ q_1=iC_1 r;\ \ C_0,\ C_1\ {\rm real},
$$
$$
q_{2n}={\rm 2n\!\!-\!\!th\ real\ polynomials,}
$$
$$
q_{2n+1}=i\times ({\rm 2n+1\!\!-\!\!th\ real\ polynomials}).
$$
For a computational convenience, we modify the long range $g/r^2$ potential
to a short range potential,
$
ge^{-\epsilon r^2}/r^2
$
 and let $\epsilon\rightarrow0$ at the end.
We see that some of the local fields decouple.
This complete set is constructed by Schmidt orthogonalization method
starting from a set of functions $\{1,r,r^2,\cdots\}$.
With the above normalization of $q_n(r)$, the mass terms become
diagonal forms of the local fields $a_n(x)$, and the effective action
is written as,

$$
S(U,\psi,\psi^\dagger)
=\int dx_1[(\frac{1}{q_0}+a_0(x))^2+a_1(x)^2+a_2(x)^2+\cdots
$$
\begin{equation}
+(\frac{1}{q_0}+a_0(x))\psi^\dagger(x)F_0(p)\psi(x)
+\sum_{l=1} a_l(x)\psi^\dagger(x)F_l(p)\psi(x)
\end{equation}
$$
+\sum_{n=0} q_n(0)a_n(x)C_0(p)\sum_{m=0} q_m(0)a_m(x)$$
$$
+2\{\frac{C_0(p)}{U_0}\sum_{l=0} q_l(0)a_l(x)\}
\psi^\dagger(x)\psi(x)
-\psi^\dagger(x)\frac{\partial}{\partial x_0}\psi(x)
+\psi^\dagger(x)\frac{p^2}{2m}\psi(x)
+\mu\psi^\dagger(x)\psi(x)
$$

The coupling strengths of local fields are given by
$$
F_l(p)=-2\int dr\{U_0(r)e^{ipr}q_l(r)
-U_0(0)q_l(0)\}\frac{V(r)}{4},
$$
\begin{equation}
C_0(p)=\int dr\vert U_0(0)\vert^2\frac{V(r)}{4}(e^{ipr}-1).
\end{equation}
The low energy and long distance physics are determined by $F_l(p_F)$.
They behave for $l\ge2$, as
\begin{equation}
F_l(p_F)=const\ \epsilon^{\frac{1}{4}}
\qquad(l\ge2),
\end{equation}
when the modified form of the potential is used.
Obviously $F_l(p_F)\ (l\ge2)$ vanish in the small $\epsilon$ limit,
and $a_l(x)\ (l\ge2)$ decouple.

Hereafter, we take only $a_0(x)$ and $a_1(x)$ into
account and study the system with them.
The action then is given by
\begin{equation}
S(a_0,a_1,\psi,\psi^\dagger)
=\int dx[(\frac{1}{q_0}+a_0(x))^2+a_1(x)^2
\end{equation}
$$
+q_0^2a_0(x)C_0(p)a_0(x)
+(\frac{1}{q_0}+a_0(x))\psi^\dagger(x)F_0(p)\psi(x)+
a_1(x)\psi^\dagger(x)F_1(p)\psi(x)
$$
$$
+2\{\frac{C_0(p)}{U_0}q_0a_0(x)\}\psi^\dagger(x)\psi(x)
-\psi^\dagger(x)\frac{\partial}{\partial x_0}\psi(x)
+\psi^\dagger(x)\frac{p^2}{2m}\psi(x)+\mu\psi^\dagger(x)\psi(x)]
$$
We consider fluctuations of the fermion field of momentum
near the Fermi momenta $\pm p_F$, and use the
following linearization approximation :
\begin{equation}
F_0(p)=F_0(\pm p_F)+(p\mp p_F)F'_0(\pm p_F),
\end{equation}
\begin{eqnarray*}
&a_0(x)F_0(p)=a_0(x)F_0(\pm p_F),\\
&\frac{p^2}{2m}=\frac{p_F^2}{2m}+(p\pm p_F)\frac{\pm p_F}{m}.
\end{eqnarray*}

The chemical potential makes
single-body energy vanishes at the Fermi momenta $\pm p_F$.

It is convenient to introduce chiral fields $\psi_L$ and
$\psi_R$ that have a momentum near $p_F$ or a momentum near
$-p_F$.
Two chiral fields are further
combined into a two component Dirac field
and the action is written as
\begin{equation}
S=C\int d\bar x[\frac{1}{v_0^2}\{\frac{1}{1+2C_0(p)q_0/(U_0F_0(p_F))}
\bar a_0(x)\}^2+\frac{1}{v_1^2}
\bar a_1^2(x)+q_0^2a_0(x)C_0(p)a_0(x)
\end{equation}
$$
+\bar{\psi}\gamma_0(p_0+\bar{a_0})\psi
+\bar{\psi}\gamma_1(\bar{p_1}+\bar{a_1})\psi],
$$
where
$$
\begin{array}{c l}
\bar a_0(x)&=\{F_0(p_F)+2\frac{C_0(p)q_0}{U_0}\}a_0(x), \\
\bar a_1(x)&=F_1(p_F)a_1(x), \\
\end{array}
$$
$$
\bar p_1=p_1C,\
\gamma_0=i\left(
\begin{array}{c c}0&1\\1&0\\
\end{array}\right),\
\gamma_1=\left(
\begin{array}{c c}0&1\\-1&0\\
\end{array}\right),\
\psi=\left(
\begin{array}{c}\psi_L\\ \psi_R\\
\end{array}\right),\
\bar\psi=\psi^\dagger i\gamma_0,
$$
$$
v_0=F_0(p_F),\ v_1=F_1(p_F),\
d\bar x=dx_0dx_1C^{-1},\ C=\frac{F_0'(p_F)}{q_0}+\frac{p_F}{m}.
$$
The action is reduced, further to
$$
S_{eff}=\int d\bar x[\frac{C}{v_0^2}(\bar a_0(x)^2+r^2\bar a_1(x)^2)
+\frac{1}{4\pi}\bar f_{\mu\nu}\frac{1}{\bar \partial^2}
\bar f_{\mu\nu}
+{\rm higher\ derivatives}]
$$
\begin{equation}
=\int d\bar x[C\frac{1}{v_0^2}\{(\partial_0\lambda)^2+r^2(\bar\partial_1
\lambda)^2+r^2(\partial_0\phi)^2+(\bar\partial_1\phi)^2
+2(1-r^2)\partial_0\lambda\bar\partial_1\phi\}
+\frac{1}{2\pi}(\bar\partial_\mu\phi)^2
\end{equation}
$$
\qquad\qquad\qquad+\frac{1}{2\pi}(\bar\partial_\mu\phi)^2
+{\rm higher\ derivatives}],
$$
$$
\bar f_{\mu\nu}=\bar \partial_\mu \bar a_\nu
-\bar \partial_\nu \bar a_\mu,\
\bar \partial^2=\partial_0^2+\bar\partial_1^2,\
\bar a_\mu=\bar \partial_\mu\lambda(\bar x)+\epsilon_{\mu\nu}
\bar\partial_\nu\phi(\bar x),\ r^2=v_0^2/v_1^2,
$$
and the inverse of the Dirac operator is written as,
\begin{equation}
\langle x\vert[\gamma_\mu(\bar p^\mu+\bar a^\mu)]^{-1}
\vert y\rangle=e^{i\lambda(\bar x)-\gamma_5\phi(\bar x)}
S_0(x-y)e^{-i\lambda(\bar y)-\gamma_5\phi(\bar y)},
\end{equation}
$$
S_0(x-y)=\langle x\vert[\gamma_\mu\bar p^\mu]^{-1}\vert y\rangle
,\ \gamma_5=-i\gamma_0\gamma_1=
\left(
\begin{array}{c c}-1&0\\0&1\\
\end{array}\right).
$$
Combining Eq.(21) and Eq.(22), we have
$$
\langle\psi(x)\psi^\dagger(y)\rangle_{x_0=y_0}
=\frac{1}{N}\int d\lambda d\phi e^{-S_{eff}(\lambda,\phi)}
e^{i(\lambda(\bar x)-\lambda(\bar y))-
\gamma_5(\phi(\bar x)-\phi(\bar y))}\times\qquad\qquad
$$
$$
\qquad\qquad\{\cos p_F(y-x){\rm Tr}[\gamma_0S_0(x-y)]
+\sin p_F(y-x){\rm Tr}[\gamma_1S_0(x-y)]\}
$$
\begin{equation}
=\frac{\sin p_F\vert y-x\vert}{\pi (x-y)^{\beta_F}}
,\
 \beta_F=1-\frac{N}{4\pi}[\frac{\xi-1}{\xi(\xi+1)}(\frac{C}{v_1^2}-
\frac{C}{v_0^2})-\frac{1}{2\pi\xi}],
\end{equation}
$$
\langle\rho(x)\rho(y)\rangle_{x_0=y_0}
=const+{\rm Tr}(\gamma_0S_0(x-y)\gamma_0S_0(y-x))
\qquad\qquad\qquad\qquad\qquad
$$
$$
+\frac{1}{N}\int d\lambda d\phi e^{-S_{eff}(\lambda,\phi)-
2\gamma_5(\phi(x)-\phi(y))}S_0(x-y)S_0(y-x)\cos 2p_F(x-y)
$$
\begin{equation}
=const+\frac{1}{2\pi^2(x-y)^2}-\frac{\cos2p_F(x-y)}{2\pi^2(x-y)^\alpha}
,
\end{equation}
$$
\alpha=2-\frac{N}{\pi}[\frac{C}{\xi v_1^2}+\frac{1}{\xi(\xi+1)}
(\frac{C}{v_0^2}-\frac{C}{v_1^2})],
$$
where
$$
N=\frac{v_1^2}{C}\frac{1}{C/v_0^2+1/2\pi},\ \xi=\sqrt{
\frac{v_1^2+2\pi C}{v_0^2+2\pi C}}.
$$
Exponents $\alpha$ and $\beta_F$ satisfy Luttinger liquid relations\cite{B}
but their
$g$ dependence are different from those obtained by Kawakami and Yang \cite{A}
based on conformal field theory.
The reason for this $g$ dependence is due to the coupling dependent kinetic
term, $F(p_1)$, in Eq.(10) and (20). This term does not exist and $\beta_F$
goes to infinity as $g$ goes to infinity in short range potential models.

{\it (B) Self-consistent stationary phase approximation} ---
In the previous part, we see that the mean field $U_0(x_1,y_1)$ of
Eq.(8) is a solution of classical equation of motion, Eq.(7).
However, it does not make the total free energy stationary due to
the higher order correction as is seen in Eq.(23).
Especially the long distance behavior of propagator is very different
from the free propagator. In this section we find a self-consistent
mean field that includes the long distance fluctuations, and
we compute current correlation functions under the self-consistent
mean field. The mean field now satisfies
\begin{equation}
\int {\cal D}U\{U-\langle\psi^\dagger\psi\rangle_U\}
e^{-S_{eff}(U)}=0,
\end{equation}
$$
\langle U\rangle=\tilde U_0.
$$

We start from a modified bi-local mean field $\tilde U_0$ which
incorpolates the radiative correction in the long distance region
and compute the correction to the propagator.
With a suitable $\tilde U_0$, the full propagator agrees to the
initial $\tilde U_0$ and full self-consistency condition is satisfied.
We find such $\tilde U_0$ first and compute other quantities next.

We start from the following form of the $\tilde U_0$:
\begin{equation}
\tilde U_0(r)=
\left\{
\begin{array}{lr}
\frac{\sin p_Fr}{\pi r}(\frac{r}{r_0})^{1-\beta_F},&\ r_0<\vert r\vert\\
\frac{\sin p_Fr}{\pi r},&\ \vert r\vert\leq r_0\\
\end{array}
\right.
\end{equation}
which has a fractional power in the long distance region and normalized
to the density $\rho$ at the origin, and substitute it to Eq.(16).
A parameter $\beta_F$ is unknown in the beginning and is determined from
the full self-consistency condition. Another parameter $r_0$ is
connected with a short distance dynamics but is regarded as a constant
in this paper.

We repeat the same procedure as before. Namely we first expand the
bi-local field with a new complete set of functions $\{\tilde q_n(r)\}$
which satisfies,
\begin{equation}
\int dr\frac{V(r)}{4}[\vert \tilde U_0(r)\vert^2\tilde q_n(r)\tilde q_m(-r)
-\tilde U_0(0)^2\tilde q_n(0)\tilde q_m(0)]
=\delta_{n,m}
\end{equation}
as
$$
U(x_1,y_1;x_0)=\tilde U_0(x_1-y_1)[1+
\sum_{n=0}^{\infty}\tilde a_n(\frac{x_1+y_1}{2};x_0)
\tilde q_n(x_1-y_1)].
$$
The action is expressed with local fields $\tilde a_n$ in the same manner
as Eq.(15) with couplings $F_l(p)$ obtained by replacing $U_0$ with
$\tilde U_0$. $F_l(p_F)$ behaves as,
\begin{equation}
F_l(p_F)=const\ \epsilon^{\frac{1}{4}},\ \ (l\geq\beta_F+\frac{1}{2}).
\end{equation}
Small $\epsilon$ limit of $F_l(p_F)$ depends on the exponent $\beta_F$.
$F_l(p_F)$ vanishes in the small $\epsilon$ limit if
$l\geq\beta_F+\frac{1}{2}$.
Hence the number of local fields that couple with the fermionic system
in this limit depends on the $\beta_F$.
It is necessary to treat the effective Lagrangian separately depending
on the magnitude of $\beta_F$.

(i) $1\leq\beta_F<\frac{3}{2}$ :
For this value of $\beta_F$ two local fields $\tilde a_0(x)$ and $\tilde
a_1(x)$ couple with fermion and the others decouple in the
$\epsilon\rightarrow 0$ limit. Hence the situation is the same as
$\beta_F=1$
case.

(ii) $\frac{3}{2}\leq\beta_F<2$ :
In this case three local fields $\tilde a_0(x),\
\tilde a_1(x),$ and $\tilde a_2(x)$ couple.
The interaction parts are given by,
\begin{equation}
\sum_{n=0}^{2}\int dx\psi^{\dagger}(x)F_n(p)\psi(x)\tilde a_n(x),
\end{equation}
$$
F_n(p)=2\int dr\{\tilde U_0(r)\tilde q_n(r)e^{ipr}-
\tilde U_0(0)\tilde q_n(0)\}\frac{V(r)}{4}.
$$
The couplings $F_n(p)$ are approximated with the Fermi momentum value
$F_n(\pm p_F)$ as far as the long wave length physics is concerned.
The couplings $F_n(\pm p_F)$ are even functions if $n$ is even and
are odd functions if $n$ is odd,
\begin{equation}
F_{0\atop2}(-p_F)=F_{0\atop2}(p_F),
\end{equation}
$$
F_1(-p_F)=-F_1(p_F).
$$

We find two component field representation of the interaction part as,
\begin{equation}
\bar \psi(x) \gamma_0\tilde F_0(p_F)\psi(x) b_0(x)+
\bar \psi(x) \gamma_1 F_1(p_F)\psi(x) b_1(x),
\end{equation}
$$
\tilde F_0(p_F)=\sqrt{F_0(p_F)^2+F_2(p_F)^2},
$$
$$
b_0(x)=\frac{F_0(p_F)}{\tilde F_0(p_F)}\tilde a_0(x)+
\frac{F_2(p_F)}{\tilde F_0(p_F)}\tilde a_2(x).
$$
With the other component $b_1(x)$ and $b_2(x)$ defined as
$$
b_1(x)=\tilde a_1(x),\ \
b_2(x)=-\frac{F_2(p_F)}{\tilde F_0(p_F)}\tilde a_0(x)+
\frac{F_0(p_F)}{\tilde F_0(p_F)}\tilde a_2(x),
$$
the mass term has the equivalent form as before,
$b_0^2+b_1^2+b_2^2=\tilde a_0^2+\tilde a_1^2+\tilde a_2^2$.

(iii) $2\leq\beta_F$ :
More fields couple with fermionic system. In the long distance region
interaction terms are written into the two component form as,
\begin{equation}
\bar \psi(x) \gamma_0\tilde F_0(p_F)\psi(x) b_0(x)+
\bar \psi(x) \gamma_1\tilde F_1(p_F)\psi(x) b_1(x),
\end{equation}
where,
$$
\tilde F_0(p_F)=\sqrt{\sum_{n=even}F_n^2},\ \
\tilde F_1(p_F)=\sqrt{\sum_{n=odd}F_n^2},
$$
$$
b_0(x)=\frac{1}{\tilde F_0(p_F)}\{\sum_n\tilde F_{2n}(p_F)\tilde a_{2n}(x)
\},
$$
$$
b_1(x)=\frac{1}{\tilde F_0(p_F)}\{\sum_n\tilde F_{2n+1}
(p_F)\tilde a_{2n+1}(x)\}.
$$
Thus the fields $b_0(x)$ and $b_1(x)$ couple with the fermion
in the long distance region with the effective coupling strength
$\tilde F_0(p_F)$ and $\tilde F_1(p_F)$ and the other fields decouple.
Consequently, the effective Lagrangian has the equivalent form
to that of the previous case.

Now we solve the full self-consistency condition with the effect of the
low energy and long wave fluctuations around the mean field included.
We have the self-consistency conditions, from Eq.(25),
$$
\langle b_n\rangle=0,
$$
\begin{equation}
\beta_F(output)=\beta_F(input),
\end{equation}
We find the exponent $\beta_F$ numerically
as a function of the coupling constant $g$.
The exponent $\nu=\beta_F-1$ thus obtained for $p_F r_0=1,\ 0.1$
is shown in Fig.1.

At the end we compute the ground state energy that includes corrections
from the low energy and long wave length fluctuations.
Since the partition function at the low temperature is expressed as
$\exp(-\beta E_0)$ with the ground energy $E_0$, the ground state energy
is found from the partition function.
We find the energy in the lowest order from the ${\cal Z}_0$ and its
correction from the $\tilde{\cal Z}$. We use the linerized form
of the Lagrangian. This Lagrangian is good in
$p_1\leq p_F$. The energy density is expressed as
\begin{equation}
{\cal E}_0=\frac{p_F^3}{6\pi m}-\int dr\frac{V(r)}{4}\{\tilde U_0(r)(2U_0(r)-
\tilde U_0(r))-\frac{p_F^2}{\pi^2}\},
\end{equation}
$$
\Delta{\cal E}=\frac{1}{8\pi^2}\int dp_0dp_1\log [1+
\frac{F_0^2p_0^2+F_1^2p_1^2C^2-\tilde q_0^2F_1^2C^2p_1^2C_0(p_1)}
{4\pi C(p_0^2+(Cp_1)^2)(1-\tilde q_0^2C_0(p_1))}].
$$
The expression of $\Delta{\cal E}$ is valid in the low energy region.
We specify the region of integration as $\vert p_1\vert\leq p_F,\
\vert p_0\vert\leq Cp_F$.
The mean field energy density ${\cal E}_0$ is given in Fig.2, and
the energy density ${\cal E}_0+\Delta{\cal E}$
from these regions are given in Fig.3.
In the same figure, the exact energy value obtained by Sutherland is given.
We see that our value is close to the exact value.
Thus long wave fluctuations are responsible to the most part of the ground
state energy.

As a conclusion, we see that the one dimensional system with long range
$1/r^2$ interaction can be expressed with the self-consistent
bi-local mean field theory.
The ground state energy agrees well with, but the exponents of the correlation
functions deviate from the exact value if the linearized form is used.

The present work is partially supported by a Grant-in-Aid for general
Scientific Research (03640256), and the Grant-in-Aid for Scientific
Research on Priority Area (04231101), the Ministry of Education,
Science and Culture, Japan.


{\bf \Large Figure Captions}

\

\noindent
Fig.1 : The exponent $\nu=\beta_F-1$ is given as a function of the coupling
constant. The dash line shows the value obtained from conformal field theory
and dotts show the self-consistent mean field values for $p_Fr_0=0.1$ and
$p_Fr_0=1.0$. The real line shows the simple mean field value.

\

\noindent
Fig.2 : The ground state energy density is given.
The real line shows the simple mean field value, and the dotts show the
self-consistent mean field values for $p_Fr_0=0.1$ and
$p_Fr_0=1.0$. The dash line is the exactly known value.

\

\noindent
Fig.3 : The ground state energy density with fluctuations included are given.
The real line shows the simple mean field value, and the dotts show the
self-consistent mean field values for $p_Fr_0=0.1$ and
$p_Fr_0=1.0$. The dash line is the exactly known value.


\begin{thebibliography}{99}
\bibitem{F}B.Sutherland, J.Math.Phys.{\bf 12}(1971)246,251; \\
Phys.Rev.{\bf A4}(1971)2019,ibid,{\bf A5}(1971)1372.
\bibitem{G}F.Calogero, J.Math.Phys.{\bf 10}(1969)2191.
\bibitem{A}N.Kawakami and S.-K.Yang, Phys.Rev.Lett.{\bf 67}(1991)2493,\\
 See a review in Prog.Theor.Phys.
Suppl.{\bf 107}(1992)59.
\bibitem{C}J.M.Luttinger, J.Math.Phys.{\bf 4}(1963)1154;\\
S.Tomonaga, Prog.Theor.Phys.{\bf 5}(1950)544.
\bibitem{B}F.D.M.Haldane, Phys.Rev.Lett.{\bf 45}(1980)1358,
ibid,{\bf 47}(1981)1840.
\bibitem{D}D.C.Mattis and E.H.Lieb, J.Math.Phys.{\bf 6}(1965)304;\\
A.Luther and I.Peschel, Phys.Rev.{\bf B9}(1974)2911.
\bibitem{I}R.B.Laughlin, Phys.Rev.Lett.{\bf 50}(1983)1395
\bibitem{E}K.Ishikawa, Prog.Theor.Phys.Suppl.{\bf 107}(1992)167;
Prog.Theor.Phys.{\bf 88}(1992)881;\\ K.Ishikawa and N.Maeda,``A Mean Field
Theory for the Quantum Hall Liquid II,\\--The Vortex Solution--",
Hokkaido University preprint EPHOU 93-001.
\bibitem{H}R.L.Stratonovich, Soviet Phys. Doklady,{\bf 2}(1957)416;\\
J.Hubbard, Phys.Rev.Lett.{\bf 3}(1959)77.
\end{thebibliography}
\end{document}